\documentclass{ws-mpla}
\usepackage{url}
\usepackage[super]{cite}
\usepackage{graphicx}
\usepackage{hyperref}

\usepackage{xcolor}

\begin{document}

%%%%%%%%%%%%%%%%%%%%% Publisher's Area please ignore %%%%%%%%%%%%%%
\catchline{}{}{}{}{}
%%%%%%%%%%%%%%%%%%%%%%%%%%%%%%%%%%%%%%%%%%%%%%%%%%%%%%%%%%%%%%%%%%%

\title{Charmonium pair production at the LHCb in the ICEM using the
$k_T$--factorization: DPS versus SPS}

\author{Alexey Chernyshev}

\address{Samara National Research University, Moskovskoe Shosse,
34, 443086, Samara, Russia.\\
aachernyshoff@gmail.com}

\author{Vladimir Saleev}

\address{Samara National Research University, Moskovskoe Shosse,
34, 443086, Samara, Russia }
\address{Joint Institute for Nuclear Research, Dubna, 141980
Russia.\\
saleev.vladimir@gmail.com}

\maketitle

%\pub{Received (Day Month Year)}{Revised (Day Month Year)}

\begin{abstract}
In the article, we study pair production of $\psi[1S]$ and
$\psi[2S]$ charmonium states in the forward rapidity region of the
LHCb experiment at the $\sqrt{s} = 13$ TeV using the improved color
evaporation model and the parton Reggeization approach, which is a
gauge invariant version of the $k_T$--factorization approach. Taking
into account single and double parton scattering mechanisms and
using the set of model parameters which have been fixed early, we
compare our theoretical predictions with the recent LHCb
experimental data as well as with another theoretical predictions.
The numerical calculations are performed using the parton--level
Monte--Carlo event generator {\tt KaTie} and modified KMRW model for
unintegrated parton distribution functions. We have found absolutely
dominant role of the double parton scattering mechanism in the
charmonium pair production at the LHCb.

\end{abstract}

%\ccode{PACS Nos.: include PACS Nos.}

\section{Introduction}\label{sec:intro}

Processes of associated heavy quarkonium production at high energies
are in the spotlight of the theoretical and experimental
studies~\cite{D0:2Psi,D0:PsiUps,LHCb:PsiD,LHCb:UpsD,ATLAS:PsiZ,CMS:2Psi,
ATLAS:2Psi,LHCb:2Psi,LHCb:2Psi2023,LHCb:Psi12S,LHCb:PsiUps} since
they provide important information about hadronization mechanisms of
heavy quarks. In addition, such processes are considered as clean
indicators of the hard multipartonic interactions in the region of
relative small invariant masses of the heavy quarkonium pairs in
compare with jet pair or massive boson pair production events. There
is a number of different theoretical models of the heavy quarkonium
hadronization. The first one is the Color Singlet Model
(CSM)~\cite{CSM1,CSM2}, which uses nonrelativistic approximation
since relative velocity of quarks into a heavy quarkonium is
estimated as small. Complete perturbation theory in the relative
quark velocity is formulated in the nonrelativistic quantum
chromodynamics (NRQCD) approach~\cite{NRQCD}. CSM is a lowest order
approximation of the NRQCD in series of $v$: ${\cal O}\left( v^0
\right)$. Nowadays, for the purpose of the phenomenological study
the octet corrections up to ${\cal O}\left( v^2 \right)$ are
calculated in the NRQCD. Since for the charmonia $v^2 \sim 0.3$ is
not enough small for fast convergence of the perturbation series,
relativistic corrections is found to be large~\cite{He:2024}. The
alternative approach, so called, the Color Evaporation Model
(CEM)~\cite{CEM1,CEM2} is based on the quark--hadron duality
hypothesis. The improvement of the CEM has been suggested by Ma and
Vogt in the Ref.~\cite{ICEM2016} considering soft gluon emissions in
the center of mass energy of $c \bar c-$pair. Previously, the ICEM
was applied to the inclusive heavy quarkonium production including
the polarized quarkonium production in the collinear parton model
(CPM)~\cite{ICEM2021} and in the $k_T$--factorization
approach~\cite{ICEM2018,ICEM2019}, as well as to different processes
of associated heavy quarkonium production in the
$k_T$--factorization
approach~\cite{CS:2Psi,CS:PsiUps,CS:PsiZW,CS:QD}.

At high energies, where the total collision energy march large a
typical hard scale of the process, $\sqrt{s} \gg \mu$, the
$k_T$--factorization or the High Energy Factorization (HEF) may be
applied, instead of fixed order CPM calculations.  We use the Parton
Reggeization Approach (PRA) which is a gauge invariant version of
the $k_T$--factorization approach based on the high-energy Regge
limit of the quantum chromodynamics (QCD)~~\cite{Gribov,KT1,KT2},
the Lipatov Effective Field Theory (EFT)~\cite{Lipatov:1,Lipatov:2},
and the modified Kimber--Martin--Ryskin--Watt (KMRW)~\cite{KMR,MRW}
model for unintegrated parton distribution functions
(uPDFs)~\cite{NefedovSaleev2020}.

Recently, the LHCb Collaboration published  new data for pair
production of $\psi[1S]$-- and $\psi[2S]$--charmonia states at the
$\sqrt{s} = 13$ TeV in the forward rapidity
region~\cite{LHCb:2Psi2023,LHCb:Psi12S}. The data are comparing with
theoretical predictions obtained in the non-complete next-to-leading
order (NLO${}^\star$)  approximation of the CPM using the
CSM~\cite{Lansberg:2013} as well as in the PRA using the
NRQCD~\cite{He:2019}. Both theoretical predictions include
contribution only from the single parton scattering (SPS) mechanism
and demonstrate a rough agreement with LHCb data. However, it was
discovered previously that in processes of heavy quarkonium pair
production the role of double parton scattering (DPS) may be
significant, for example, see
Refs~\cite{D0:PsiUps,LHCb:PsiD,LHCb:UpsD,LHCb:PsiUps}.

In contrast with theoretical studies in
Ref.~\cite{Lansberg:2013,He:2019}, we use in current study the ICEM
as a hadronization model instead of the CSM or the NRQCD, and take
into account contributions both the SPS and DPS associated heavy
quarkonium production mechanisms.

\section{Parton Reggeization Approach}\label{sec:PRA}

For the description of hard processes at the energies of the LHC one
may use the $k_T$--factorization approach which takes into account
the Reggeization of hard scattering amplitudes and non-collinear
parton dynamics in uPDFs beyond the CPM evolution equations. In the
PRA~\cite{NSS2013,Karpishkov:2017,NefedovSaleev2020}, hadronic cross
section is related to the hard scattering coefficient by the
factorization formula proven at leading logarithmic
approximation~\cite{KT1,KT2}:
$$
d \sigma \simeq \int \frac{dx_1}{x_1} \, \int \frac{d^2 {\bf
q}_{T_1}}{\pi} \, \Phi(x_1, t_1, \mu^2) \, \int \frac{dx_2}{x_2} \,
\int \frac{d^2 {\bf q}_{T_2}}{\pi} \, \Phi(x_2, t_2, \mu^2) \times
d\hat\sigma_{\rm PRA},
$$
where $t_i = {\bf q}_{T_i}^2$, the hard scattering coefficient
$d\hat\sigma_{\rm PRA}$ is expressed via the squared Reggeized
amplitude $\overline{\mid {\cal A}_{\rm PRA} \mid^2}$ in a standard
way. Off--shell initial gluons and quarks are treated as Reggeized
partons of the gauge invariant EFT for processes in multi--Regge
kinematics proposed by L.N.~Lipatov~\cite{Lipatov:1,Lipatov:2}.

In the PRA we use the modified KMRW model~\cite{KMR,MRW} for uPDFs
and calculate the uPDF by the
formula~\cite{KMR,MRW,NefedovSaleev2020}:
$$
\Phi_i(x, t, \mu^2) = \frac{\alpha_S(t)}{2 \pi} \frac{T_i(x,t,
\mu^2)}{t} \ \sum\limits_j \int_x^1 dz \, P_{i j}(z) \,
F_j\left( \frac{x}{z}, t \right) \times \theta\left( \Delta(t, \mu)
- z \right),
$$
where $F_i(x, \mu^2) = x \, f_i(x, \mu^2)$ and $ \Delta(t, \mu) =
\mu / \left( \sqrt{t} + \mu \right)$ is a KMR--cutoff function which
ensures rapidity ordering between last emitted parton in the parton
cascade and the particles produced in the hard subprocess, and
cutoff infrared divergence, where $z \to 1$. Here and below we put
factorization and renormalization scales are equal: $\mu_F = \mu_R =
\mu$. To resolve collinear divergence problem, we require that the
modified KMRW uPDF satisfies the exact normalization condition for
the arbitrary $x$ and $t$:
\begin{equation}
\int^{\mu^2}_0 dt \, \Phi_i(x, t, \mu^2) = F_i(x, \mu^2)
\end{equation}
which can be achieved by introducing the Sudakov form--factor
$T_i(x, t, \mu^2)$, with boundary conditions $T_i(x, 0, \mu^2)=0$
and $T_i(x, \mu^2, \mu^2)=1$, as follows
\begin{equation}
\Phi_i(x, t, \mu^2) =
\frac{d}{dt}\left[ T_i(x, t, \mu^2) \, F_i(x, t) \right].
\end{equation}
Explicit solution for Sudakov form--factor, which is dependent on
$x$ opposite the KMRW model, was first obtained
in~\cite{NefedovSaleev2020}.

Previously, the PRA was successfully  used for the different
quarkonium production studies: prompt $J/\psi$ production at
TEVATRON and LHC energies~\cite{SaleevVasin2006,NSS2012}, production
of polarized heavy quarkonium states~\cite{NS2016}, pair $J/\psi$
production with high--energy resummation~\cite{He:2019} etc. The PRA
beyond the leading order (LO) in $\alpha_S$ was considered in
Refs.~\cite{PRA-NLO1,PRA-NLO2}.

\section{Improved Color Evaporation Model}\label{sec:ICEM}

Current status of the ICEM can be found in Ref.~\cite{ICEM2016}.
Since initial partons in the PRA have non--negligible transverse
momenta, a description of the single charmonium production in the
ICEM is possible already in the LO of ${\cal O}(\alpha_S^2)$ parton
subprocesses:
\begin{eqnarray}
R + R & \to & c + \bar c, \label{eq:proc1} \\
Q_q + \bar Q_q & \to & c + \bar c, \label{eq:proc2}
\end{eqnarray}
where $R$ is the Reggeized gluon and $Q_q (\bar Q_q)$ is the
Reggeized quark (antiquark), $q=u,d,s,c$. In the ICEM, final
charmonium state is treated as the $c \bar c$--pair with invariant
mass $M$ less than $D \bar D$ production threshold $2 M_D$
regardless of their quantum numbers:
\begin{eqnarray}
\frac{d\sigma^{\rm SPS}_\psi}{d^3 p_\psi} & = &
{\cal F}^\psi \times
\int_{M_\psi}^{2 M_D} dM \, d^3 {\bf p}_{c \bar c} \
\delta^{(3)}\left( {\bf p}_\psi - \frac{M_\psi}{M} {\bf p}_{c \bar c} \right) \,
\frac{d\sigma_{c\bar c}}{dM \, d^3 p_{c \bar c}}, \label{eq:ICEM}
\end{eqnarray}
where $p_{c\bar c} = p_c + p_{\bar c}$ is the $4$--momenta of a $c
\bar c$--pair, $M_\psi$ is the mass of a final charmonium. The ICEM
takes into account an important kinematical effect consisting of a
mass difference between intermediate state and final charmonium
considering soft gluon emissions~\cite{ICEM2016}: $p_\psi^\mu \simeq
\left( M_\psi / M \right) p_{c\bar c}^\mu$. The non--perturbative
factor ${\cal F}^\psi$ in the (\ref{eq:ICEM}) is considered as a
probability of hadronization of the $c \bar c$--pair with invariant
mass $M_\psi < M < 2 M_D$ into the charmonium, so the each factor is
unique for the each charmonium state $\psi [nS]$.

LO ${\cal O}(\alpha_S^4)$ partonic subprocesses in the pair
charmonium production $p + p \to \psi_1 + \psi_2 + X$ with $\psi_1 =
\psi[n_1S]$ and $\psi_2 = \psi[n_2S]$ are follows:
\begin{eqnarray*}
R + R & \to & c + \bar c + c + \bar c, \\
Q_q + \bar Q_q & \to & c + \bar c + c + \bar c.
\end{eqnarray*}
Following the ICEM, cross section for the pair charmonium production
is expressed in terms of the hadronization probability ${\cal
F}^{\psi_1\psi_2}$ of two $c\bar c-$pairs with $M_{\psi_1,\psi_2} <
M_{1, 2} < 2 M_D$ into the charmonium pair. According to the
identity principle of quantum mechanics, factor ${\cal F}^{\psi_1
\psi_2}$ is factorized into inclusive probabilities ${\cal
F}^{\psi_1} \times {\cal F}^{\psi_2}$ only in case of different
charmonium states $n_1 \neq n_2$ and one should be considered as
independent from ${\cal F}^{\psi_1}$ or ${\cal F}^{\psi_2}$ in case
of identical charmonium states since all $c \bar c$--pairs are
identical by quantum numbers in the ICEM. The cross section for the
pair charmonium production via the SPS mechanism can be written in
the following way:
\begin{eqnarray}
d \sigma^{\rm SPS}_{\psi_1\psi_2} & = &
{\cal F}^{\psi_1\psi_2} \times
\int_{M_{\psi_1}}^{2 M_D} dM_1 \, \int_{M_{\psi_2}}^{2 M_D} dM_2 \,
\frac{d\sigma_{(c\bar c)_1 (c \bar c)_2}^{\rm SPS}}{dM_1 \,
dM_2}.\label{eq:SPS}
\end{eqnarray}

Another mechanism of the pair charmonium production is the DPS.
Using the DPS "pocket--formula" obtained under the assumption of
independence of two hard parton subprocesses, cross sections are
factorized as follows:
\begin{equation}
d \sigma^{\rm DPS}_{\psi_1\psi_2}  =  \frac{1}{1 + \delta_{{n_1}
{n_2}}} \frac{d \sigma^{\rm SPS}_{\psi_1} \times d \sigma^{\rm
SPS}_{\psi_2}}{\sigma_{\rm eff}},\label{eq:DPS}
\end{equation}
where $d \sigma^{\rm SPS}_{\psi_{1,2}}$ are calculated by formula
(\ref{eq:ICEM}). Here the parameter $\sigma_{\rm eff}$  controls the
DPS contribution to the total cross section, $\delta_{{n_1}
{n_2}}$--standard Kronecker $\delta$--symbol.

Thus, the total cross section for the pair charmonium production is
the sum of the SPS~(\ref{eq:SPS}) and the DPS~(\ref{eq:DPS})
contributions. As it was shown in Ref.~\cite{Kutak:2016}, at LHC
energies both mechanisms give significant contribution to the cross
section for the production of two $c \bar c-$pairs.

\section{{\tt KaTie} event generator}\label{sec:KaTie}

The scattering processes $2 \to 4$ is a difficult task for numerical
calculations in the $k_T-$factorization based on amplitudes
originated from Feynman rules of the Lipatov EFT. For the
calculation of cross sections of such processes is more suitable to
use technic of Monte--Carlo (MC) event generators. A some time ago,
the MC parton--level generator {\tt KaTie}~\cite{KaTie} for the HEF
calculations was developed. The method used in the {\tt KaTie} for
numerical generation of the gauge--invariant amplitudes with
off--shell initial states is based on the spinor amplitudes
formalism and recurrence relations of the BFCW
type~\cite{hameren1,hameren2}. The results obtained in the {\tt
KaTie} at the stage of numerical calculations is fully equivalent to
the results obtained in the PRA for tree--level
diagrams~\cite{Kutak:2016}. {\tt KaTie} can use the {\tt TMDlib2}
library~\cite{TMDlib2}, which includes various uPDF sets, such as
most popular now, {\tt ccfm-JH}~\cite{CCFM-PDF} and {\tt
PB-NLO}~\cite{PB-PDF1,PB-PDF2} sets. Our arguments in support to use
the modified {\tt KMRW} uPDFs~\cite{NefedovSaleev2020} are given in
Ref.~\cite{CS:QD}.

\section{Results}\label{sec:res}

\boldmath
\subsection{Pair $\psi[1S] + \psi[1S]$ production}\label{sec:11S}
\unboldmath

\begin{figure}
\centering
\includegraphics[scale=0.25]{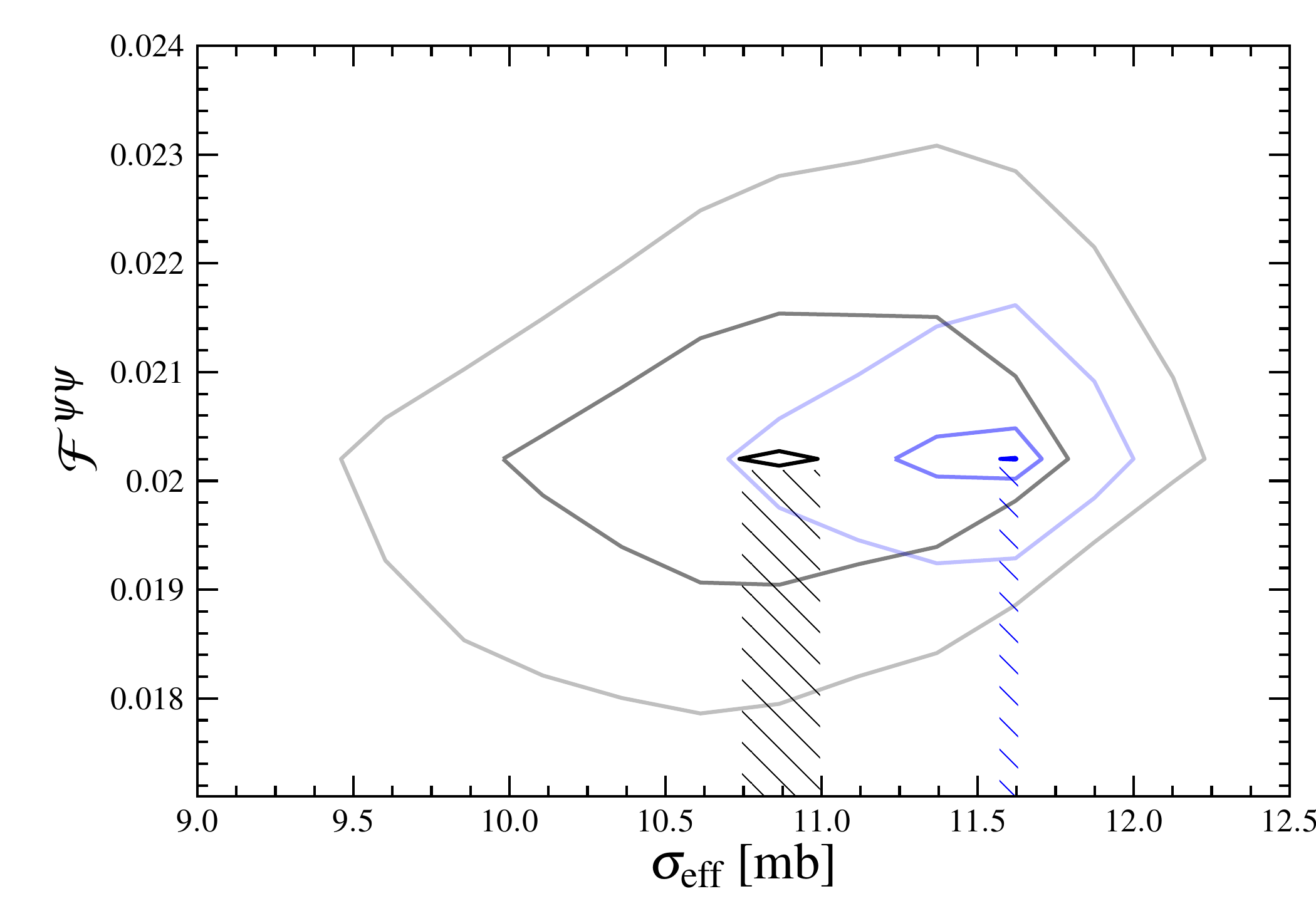}
\caption{\centering Regions of the the parameters $({\cal
F}^{\psi\psi}$ and $\sigma_{\rm eff})$ obtained by fitting data for
pair $\psi[1S]$ production: gray lines -- including
data~\cite{CMS:2Psi,ATLAS:2Psi,LHCb:2Psi}, blue lines -- including
new LHCb data~\cite{LHCb:2Psi2023}.}\label{fig:1}
\end{figure}

In Ref.~\cite{CS:2Psi}, we have fixed ICEM parameter ${\cal
F}^{\psi} \simeq 0.02$ for $\psi[1S]$--charmonium state and
performed a fit of the total cross sections for the pair $\psi[1S]$
production at high energies $\sqrt{s} = 7 - 13$ TeV by the two free
parameters $({\cal F}^{\psi\psi}$ and $\sigma_{\rm eff})$ minimizing
a function:
$$
\chi = \sum_k \frac{\mid \sigma_k^{\rm exp} - \sigma_k \mid}{\Delta
\sigma_k^{\rm exp}},
$$
where sum is taken over all experimental data. Putting $\chi = 1.0,
1.5, 2.0$, we obtained a regions of possible values of the
parameters $({\cal F}^{\psi\psi}$ and $\sigma_{\rm eff})$. The
boundaries of these regions are plotted as gray isolines in
Fig.~\ref{fig:1}. The best description of the data is reached with
${\cal F}^{\psi\psi} \simeq {\cal F}^{\psi} \simeq 0.02$ and
$\sigma_{\rm eff} \simeq 11.0$ mb. We update our previous fit by
adding new LHCb Collaboration data~\cite{LHCb:2Psi2023} at the
$\sqrt{s} = 13$ TeV in the forward rapidity region: $p_T^\psi < 14$
GeV and $2.0 < y^\psi < 4.5$ for $\psi[1S,2S]$--charmonia. New
boundaries of the parameter regions described above are plotted as a
blue isolines in Fig.~\ref{fig:1}. While condition ${\cal
F}^{\psi\psi} \simeq {\cal F}^{\psi}$ is still holding, we obtained
a new value of the DPS parameter approximately equal  to
$\sigma_{\rm eff} \simeq 11.5$ mb. In fact, these two values of the
DPS parameter, $\sigma_{\rm eff} \simeq 11.0$ mb and $11.5$ mb, are
equivalent within the theoretical uncertainties. In this work we use
previous obtained value $\sigma_{\rm eff} \simeq 11.0$ mb, which was
also used for the successful description of the data for different
processes of the associated heavy quarkonium production at high
energies, such as $J/\psi(\Upsilon)+Z(W)$~\cite{CS:PsiZW} and
$J/\psi(\Upsilon)+D$~\cite{CS:QD}.

\begin{figure}[ph]
\centering
\includegraphics[scale=0.15]{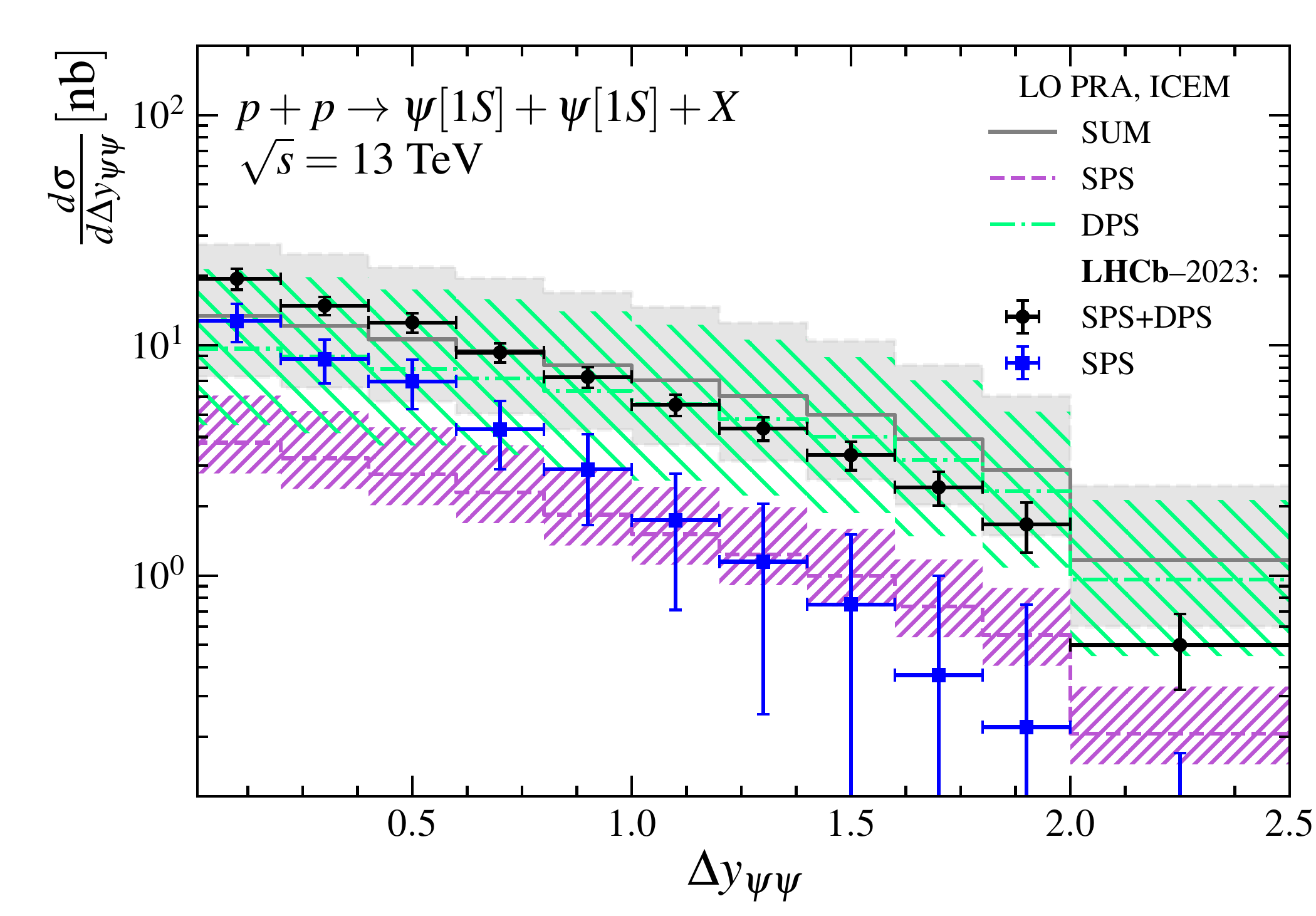}
\includegraphics[scale=0.15]{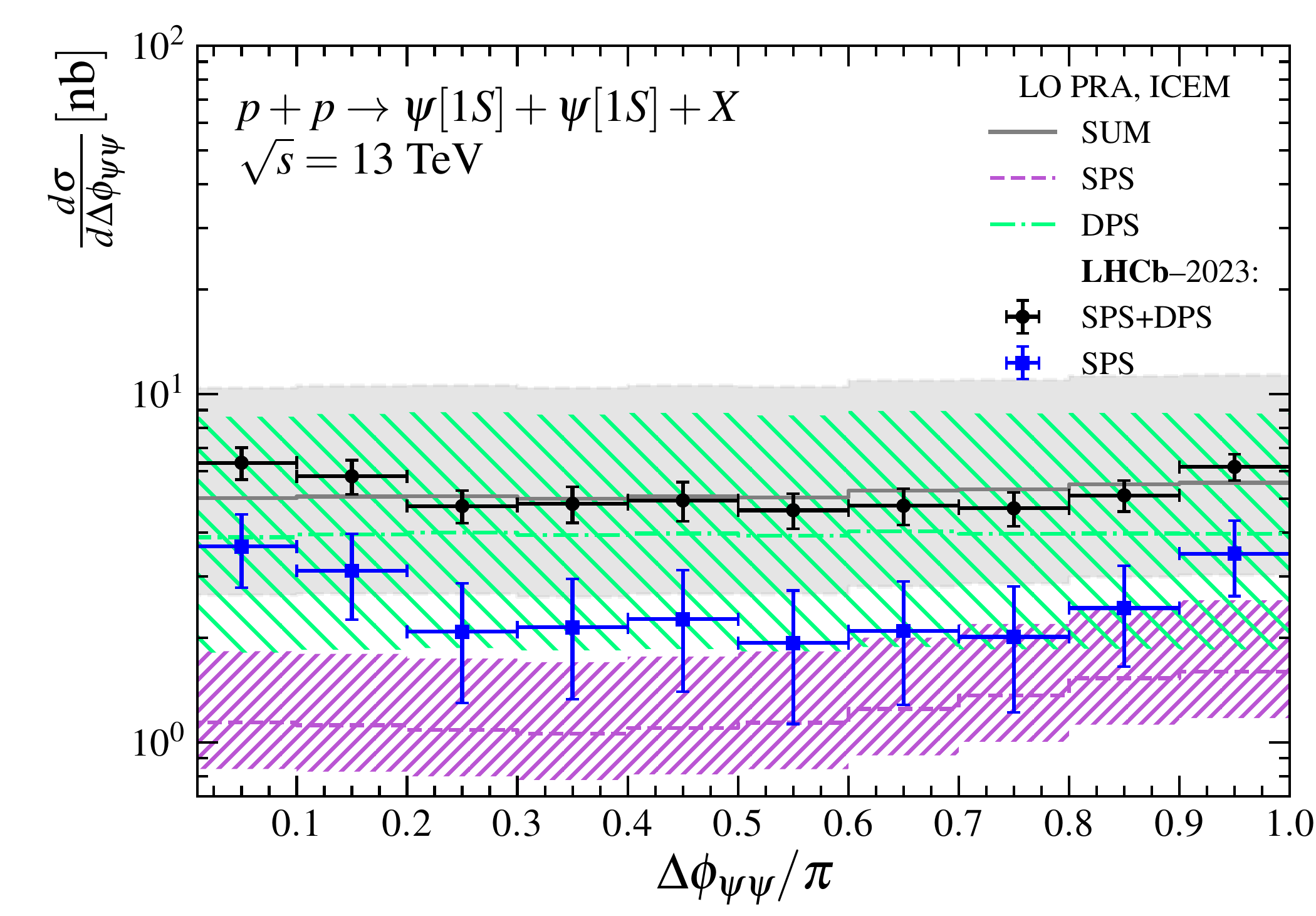}
\\
\includegraphics[scale=0.15]{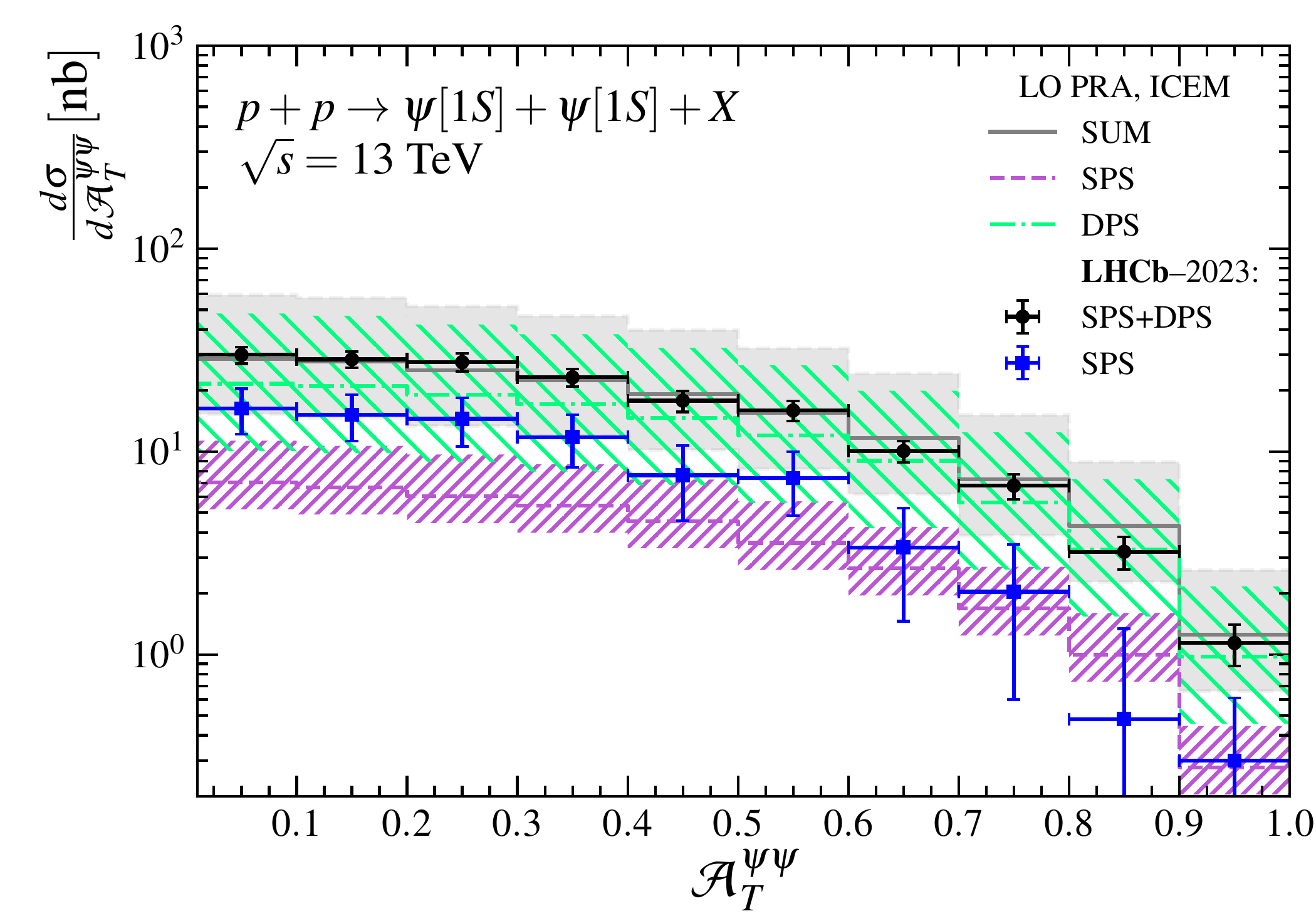}
\includegraphics[scale=0.15]{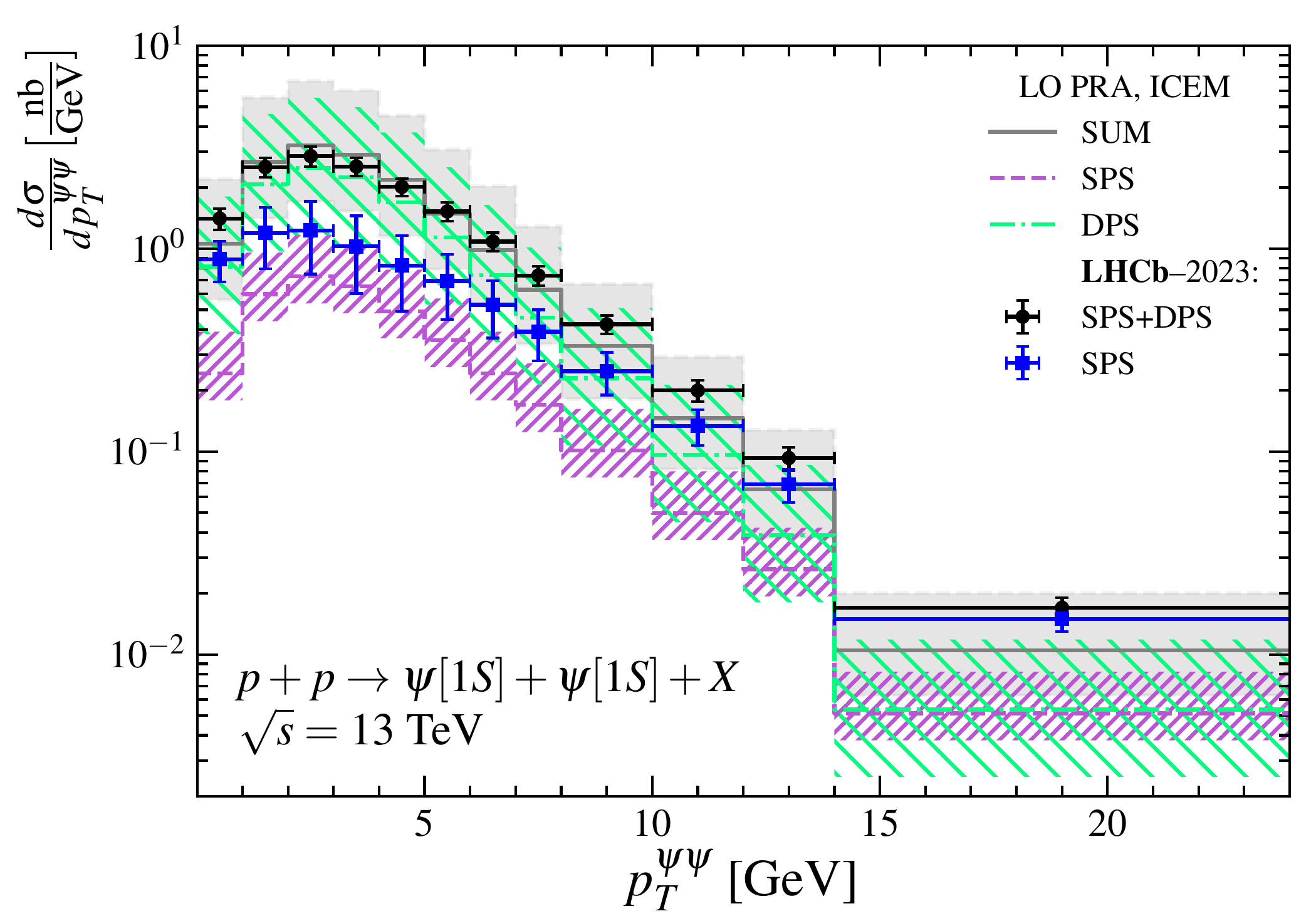}
\\
\includegraphics[scale=0.15]{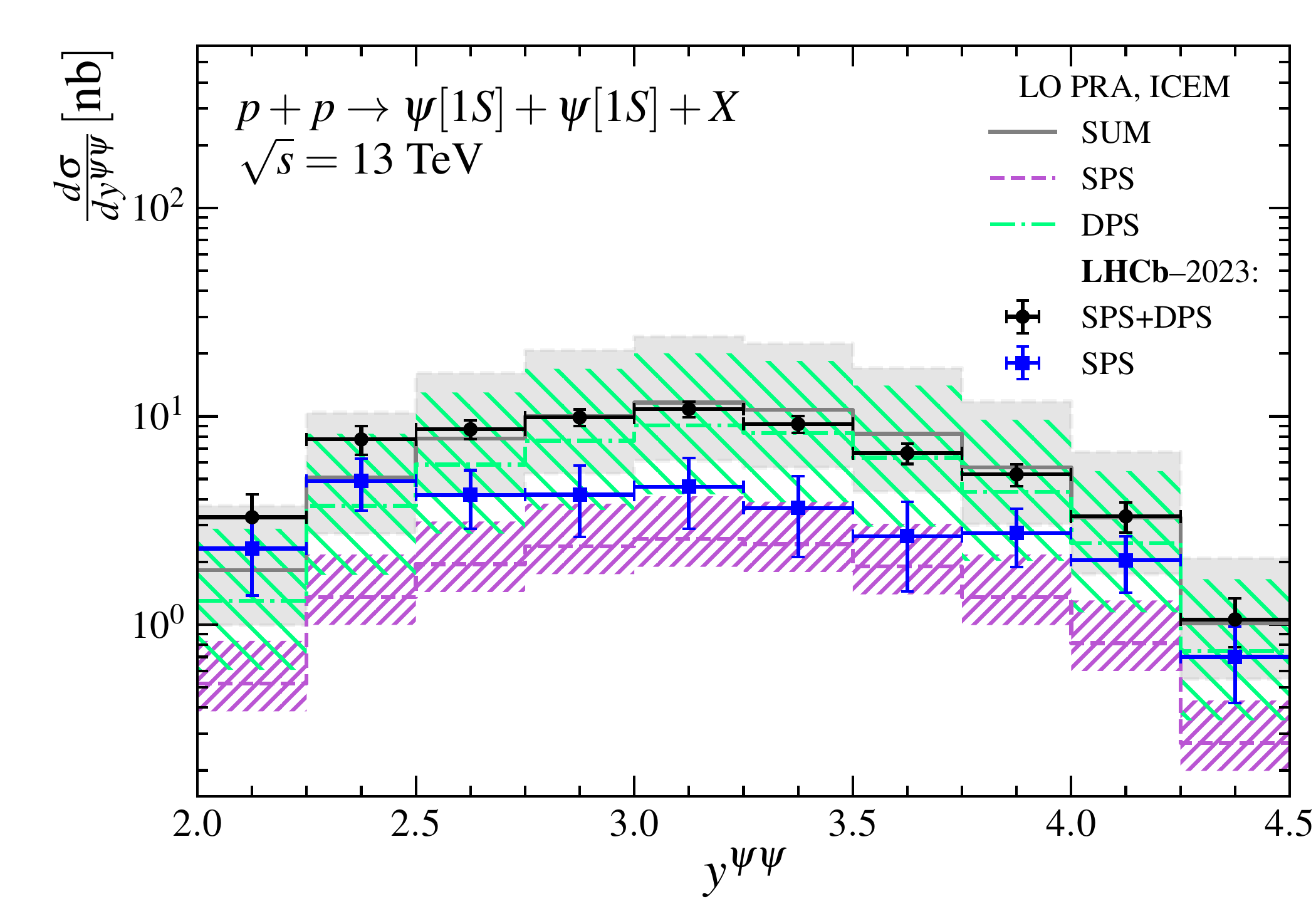}
\includegraphics[scale=0.15]{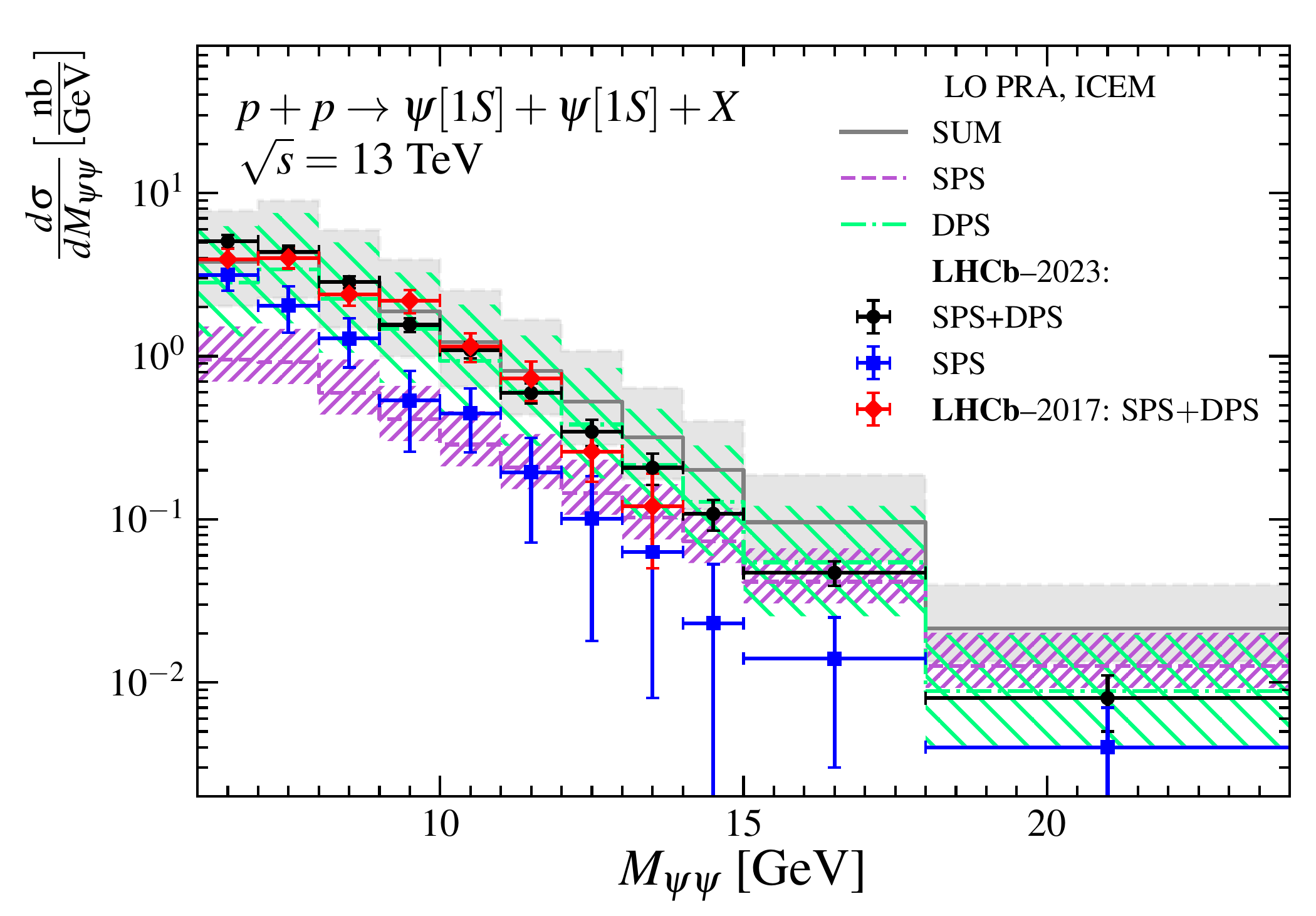}
\caption{\centering
Differential cross sections for pair $\psi[1S] + \psi[1S]$ production at
$\sqrt{s} = 13$ TeV on different kinematical variables. Contributions from SPS
and DPS mechanisms are shown separetly.
The data are from LHCb Collaboration~\cite{LHCb:2Psi2023}.}\label{fig:2}
\end{figure}

In the recent publication of the LHCb
Collaboration~\cite{LHCb:2Psi2023}, data are separated on the
contributions of the SPS and the DPS production mechanisms, see
black points in Fig.~\ref{fig:2} corresponding to the sum of the SPS
plus DPS contributions and blue points corresponding to the only SPS
contribution. Our theoretical predictions are also shown separately
for the SPS (\ref{eq:SPS}) and DPS (\ref{eq:DPS}) with the scale
uncertainties in Fig.~\ref{fig:2}. For a comparison with
experimental data, cross sections for SPS plus DPS, SPS and DPS
contributions are collected in Table~\ref{tab:1}.  We found the DPS
mechanism to be dominant so that the ratio of the SPS to DPS is
$0.31^{+0.57}_{-0.25}$. Since the predictions obtained in the PRA
using NRQCD from Ref.~\cite{He:2019} for the new LHCb Collaboration
data~\cite{LHCb:2Psi2023} are presented only for the invariant mass
spectrum, we especially  focus our discussion on this spectrum. Our
predictions for the SPS production mechanism are in good agreement
with data in the region $9<M_{\psi\psi}< 14$ GeV and underestimate
data at the $M_{\psi\psi} < 9$ GeV. Phenomenological condition
${\cal F}^{\psi\psi} \simeq {\cal F}^\psi$, which was founded early,
corresponds to the maximum possible contribution of the SPS
mechanism since ${\cal F}^{\psi\psi} \leq {\cal F}^\psi$, which
confirms our fit, see Fig.~\ref{fig:1}. Our predictions based on the
DPS mechanism are also in good agreement with data excluding region
of very large invariant mass $M_{\psi\psi}$. This disagreement may
be explained by insufficient experimental statistics at the end of
the invariant mass spectrum.  We observed such situation earlier
when describing previous LHCb Collaboration data~\cite{LHCb:2Psi},
see Ref.~\cite{CS:2Psi}.
% and red points corresponding to the sum of SPS
%and DPS contributions in Fig.~\ref{fig:2}.

\renewcommand{\arraystretch}{1.8}

\begin{table}[ht]
\tbl{Comparison of the total cross sections and ratios for pair $1S
+ 1S$ and $1S + 2S$ charmonium states production at $\sqrt{s} = 13$
TeV.} {\begin{tabular}{l | l | r r}

Process & Cross section / Ratio &
${\rm Exp.} \pm ({\rm stat.}) \pm ({\rm syst.})$ & ICEM+PRA\\

\hline

& $\sigma_{\psi[1S]\psi[1S]}^{\rm SPS + DPS}$, nb &
$16.36 \pm 0.28 \pm 0.88$ & $16.33^{+17.48}_{-7.67}$ \\

$\psi[1S]+\psi[1S]$,  & $\sigma_{\psi[1S]\psi[1S]}^{\rm SPS}$, nb &
$7.9 \pm 1.2 \pm 1.1$ & $3.9^{+2.3}_{-1.0}$ \\

LHCb~\cite{LHCb:2Psi2023} & $\sigma_{\psi[1S]\psi[1S]}^{\rm DPS}$,
nb &
$8.6 \pm 1.2 \pm 1.0$ & $12.4^{+15.1}_{-6.6}$ \\

\hline

& $\sigma_{\psi[1S]\psi[2S]}^{\rm SPS + DPS}$, nb &
$4.5 \pm 0.7 \pm 0.3$ & $5.4^{+6.7}_{-3.0}$ \\

& $\sigma_{\psi[1S]\psi[2S]}^{\rm SPS}$, nb &
-- & $\left( 2.0^{+4.0}_{-1.0} \right) \times 10^{-2}$ \\

$\psi[1S]+\psi[2S]$,  & $\sigma_{\psi[1S]\psi[2S]}^{\rm DPS}$, nb &
-- & $5.4^{+6.6}_{-3.0}$ \\

LHCb~\cite{LHCb:Psi12S} & $\sigma_{\psi[1S]\psi[2S]}^{\rm SPS +
DPS}/\sigma_{\psi[1S]\psi[1S]}^{\rm SPS + DPS}$&
$0.274 \pm 0.044 \pm 0.008$ & $0.332$ \\

& $\sigma_{\psi[1S]\psi[2S]}^{\rm
DPS}/\sigma_{\psi[1S]\psi[1S]}^{\rm DPS}$&
$0.282 \pm 0.027$ & $0.434$ \\

\end{tabular}\label{tab:1} }

\end{table}

Predictions obtained in the PRA using the NRQCD~\cite{He:2019} for
the invariant mass differential cross section, taken into account
only the SPS contribution, overestimate corresponding
data~\cite{LHCb:2Psi2023} by about one order of magnitude and
describe data for sum of the SPS and DPS contributions in the region
of high invariant mass $M_{\psi\psi} > 10$ GeV. In
Ref.~\cite{LHCb:2Psi2023} are presented predictions obtained in the
NLO${}^\star$ CPM (without loop corrections) using the CSM, see
Ref.~\cite{Lansberg:2013}, for the SPS contribution. These
predictions approximately agree with data for the sum of SPS and DPS
contributions within large theoretical uncertainties and
overestimate data for the SPS contribution by an one order of
magnitude at the upper limit of invariant mass.

The results of our calculation, obtained in the ICEM using the PRA
for different differential cross sections, are collected in
Fig.~\ref{fig:2}. We founded a rather good agreement with data for
the all spectra. For example, predicted azimuthal angle difference
$\Delta\phi_{\psi\psi}$ distribution is flat due to the dominant DPS
contribution and it does not contradict data within theoretical
uncertainties.

Taking into account satisfactory data description for the associated
heavy quarkonium production processes: $\psi[1S] +
\psi[1S]$~\cite{CS:2Psi}, $\psi[1S] +
\Upsilon[1S]$~\cite{CS:PsiUps}, $\psi[1S] + Z / W$~\cite{CS:PsiZW},
and ${\psi(\Upsilon)}[1S] + D$~\cite{CS:QD}, which we have
previously obtained, one can conclude that our choice of the ICEM
and DPS model parameters is optimal.

\boldmath
\subsection{Pair $\psi[1S] + \psi[2S]$ production}\label{sec:12S}
\unboldmath

At the first step, we have fixed parameter ${\cal F}^{\psi'}=
0.06^{+0.01}_{-0.01}$ for $\psi[2S]$--charmonium state by fitting
LHCb data~\cite{LHCb:Psi2S} for single $\psi[1S]$ production under
similar kinematical conditions taking into account contributions
from the LO subprocesses~(\ref{eq:proc1})--(\ref{eq:proc2}). The
obtained ${\cal F}^{\psi'}$  is a $3$ times larger than  ${\cal
F}^{\psi}$ for $\psi[1S]$--state. This fact can be associated with
the lower limit of integral in~(\ref{eq:ICEM}) for the $\psi[2S]$
production since $M_{\psi[2S]} > M_{\psi[1S]}$ and the integration
region becomes smaller than in case of the $\psi[1S]$ state, which
leads to reverse hierarchy of hadronization probabilities. A similar
situation was founded in the case of higher bottomonium states
production in Ref.~\cite{ICEM:2018_1}. But, as it was shown
in~\cite{ICEM2016}, ratio $\sigma_{\psi[2S]} / \sigma_{\psi[1S]}$ in
the ICEM is in a good agreement with data and the relation ${\cal
F}^{\psi'} > {\cal F}^{\psi}$ is not an artifact of this model.

Cross section for the pair $\psi[1S] + \psi[2S]$ production was
first measured by the LHCb Collaboration~\cite{LHCb:Psi12S} at
$\sqrt{s} = 13$ TeV and $p_T^{\psi} < 14$ GeV, $2.0 < y^\psi < 4.5$
for each charmonia. According to the ICEM using PRA approach
discussed above all model parameters are known before the
calculation of the pair $\psi[1S] + \psi[2S]$ production: ICEM
parameters ${\cal F}^{\psi}$ and ${\cal F}^{\psi'}$ are fixed in the
single production study, the DPS parameter $\sigma_{\rm eff}$ is
fixed in the pair $\psi[1S]$ data analysis. Thus, when considering
data~\cite{LHCb:Psi12S} for the pair production of $\psi[1S]$-- and
$\psi[2S]$--charmonium states, we present predictions without using
free parameters.

\begin{figure}[ph]
\centering
\includegraphics[scale=0.15]{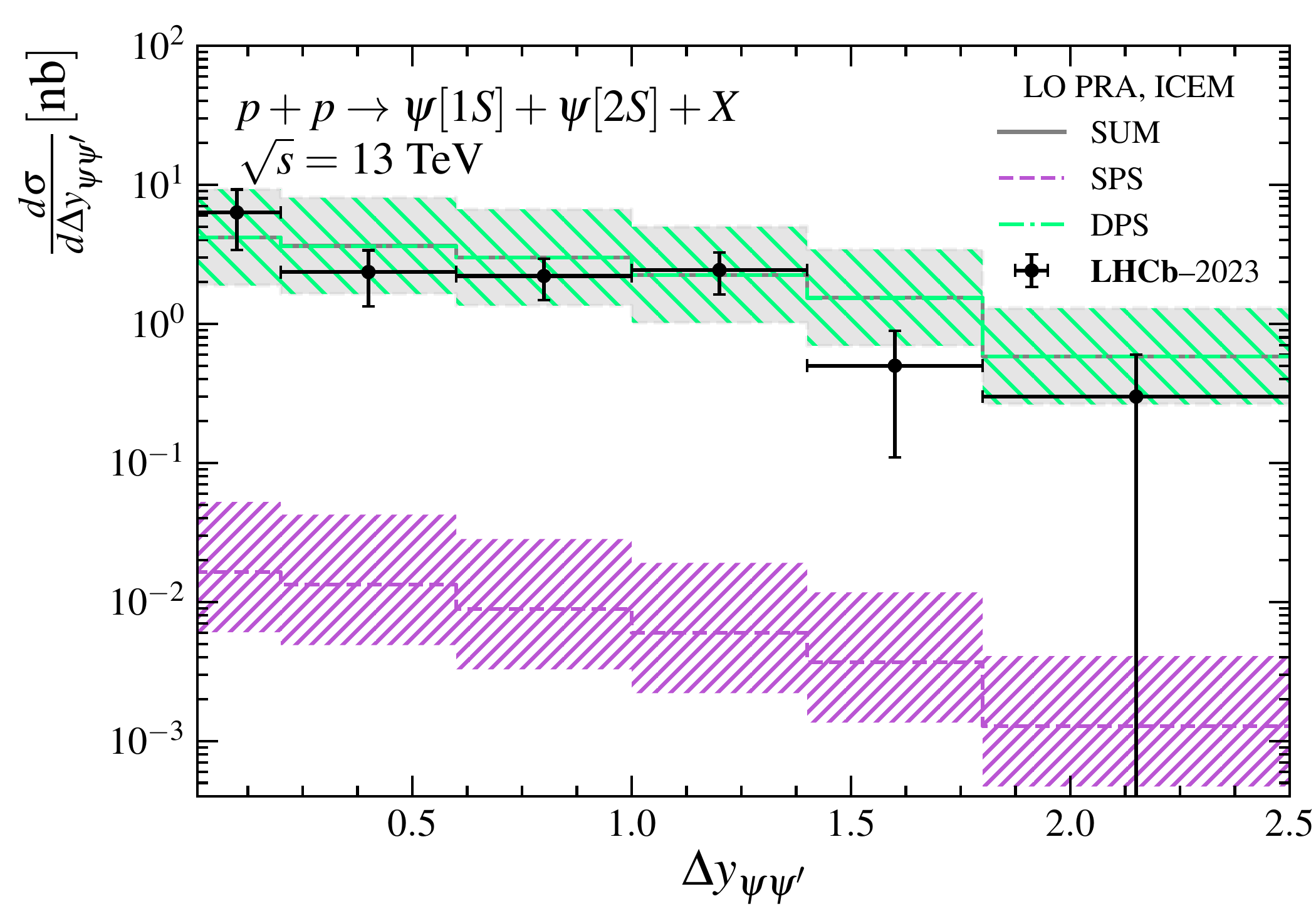}
\includegraphics[scale=0.15]{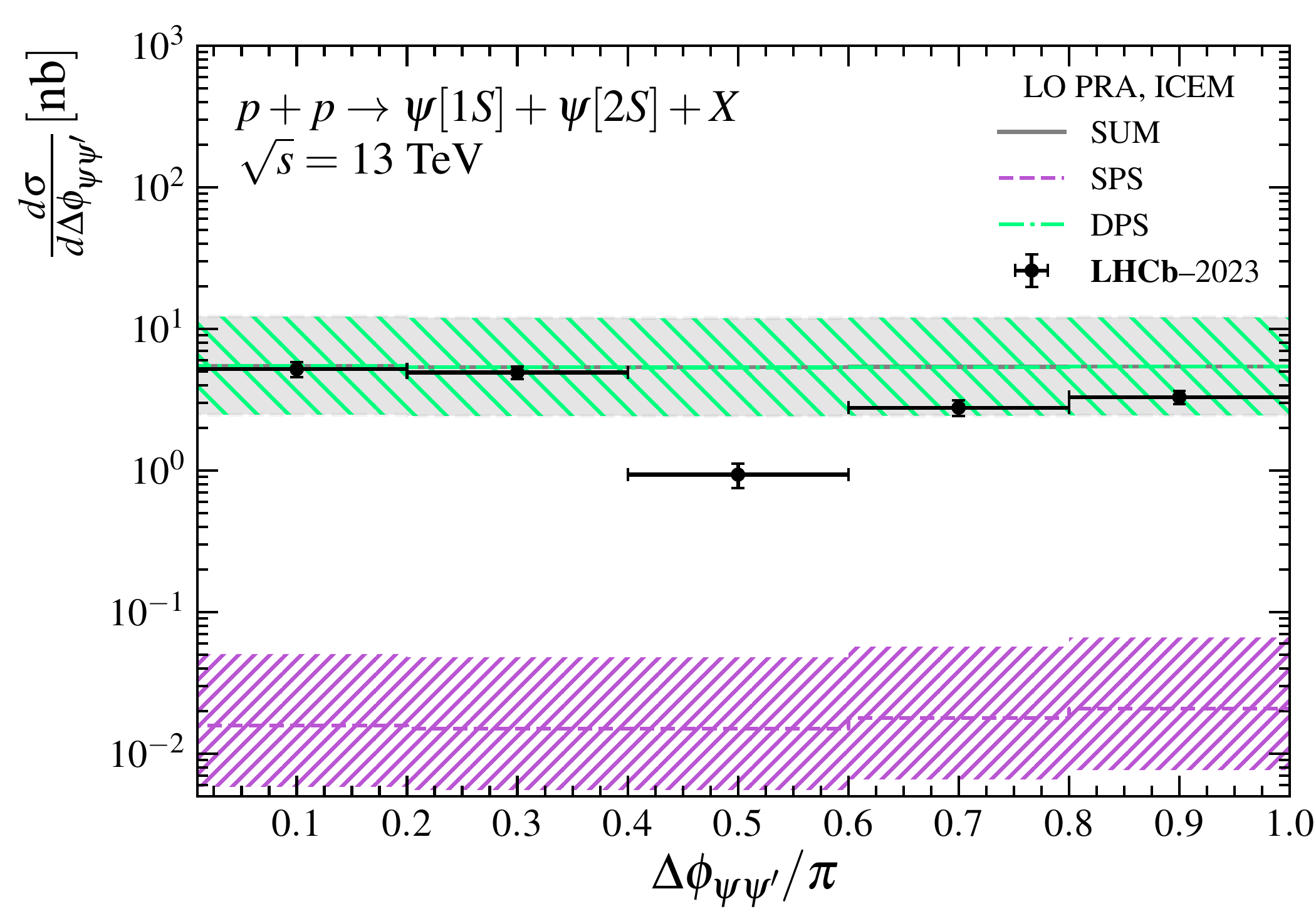}
\\
\includegraphics[scale=0.15]{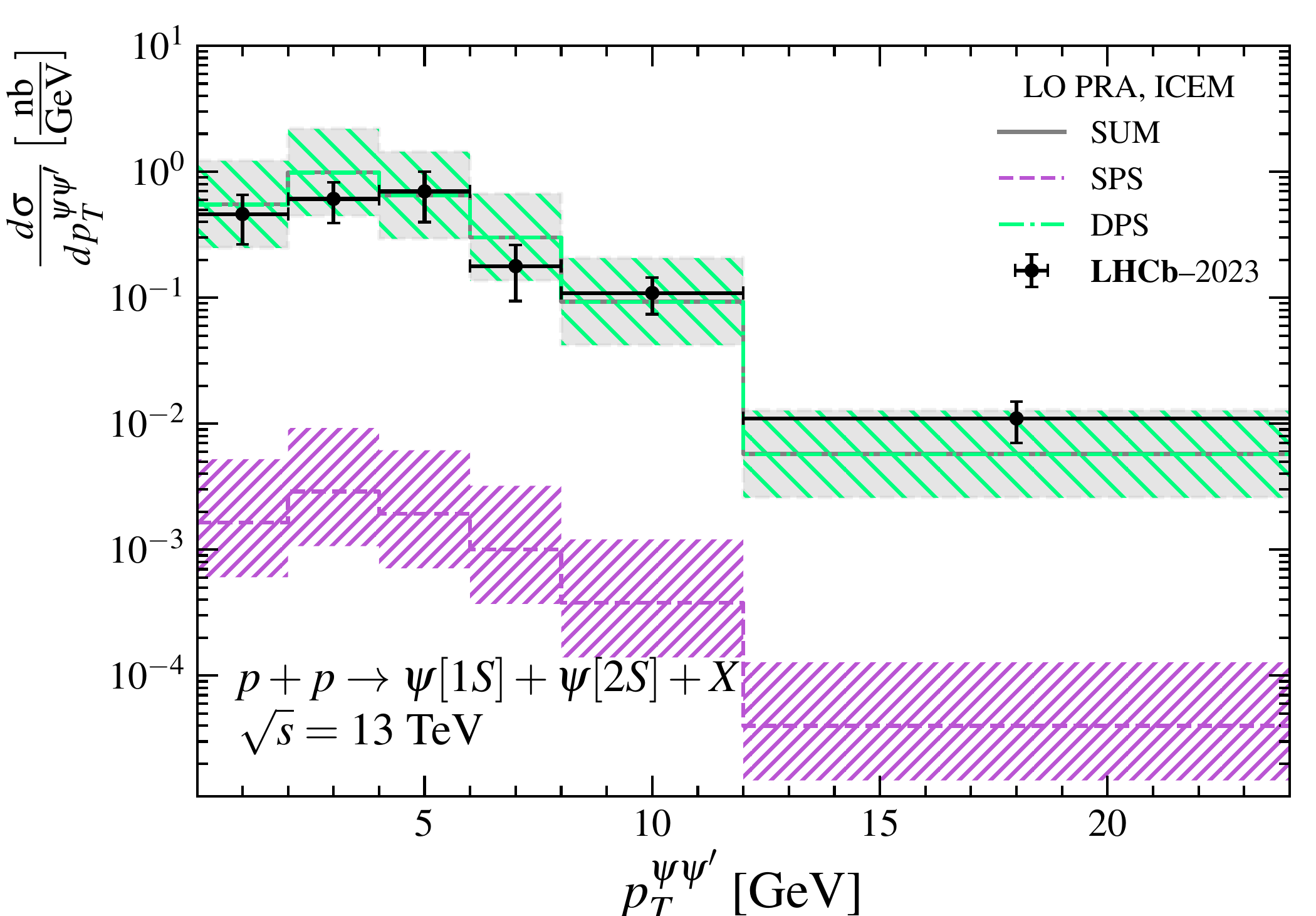}
\includegraphics[scale=0.15]{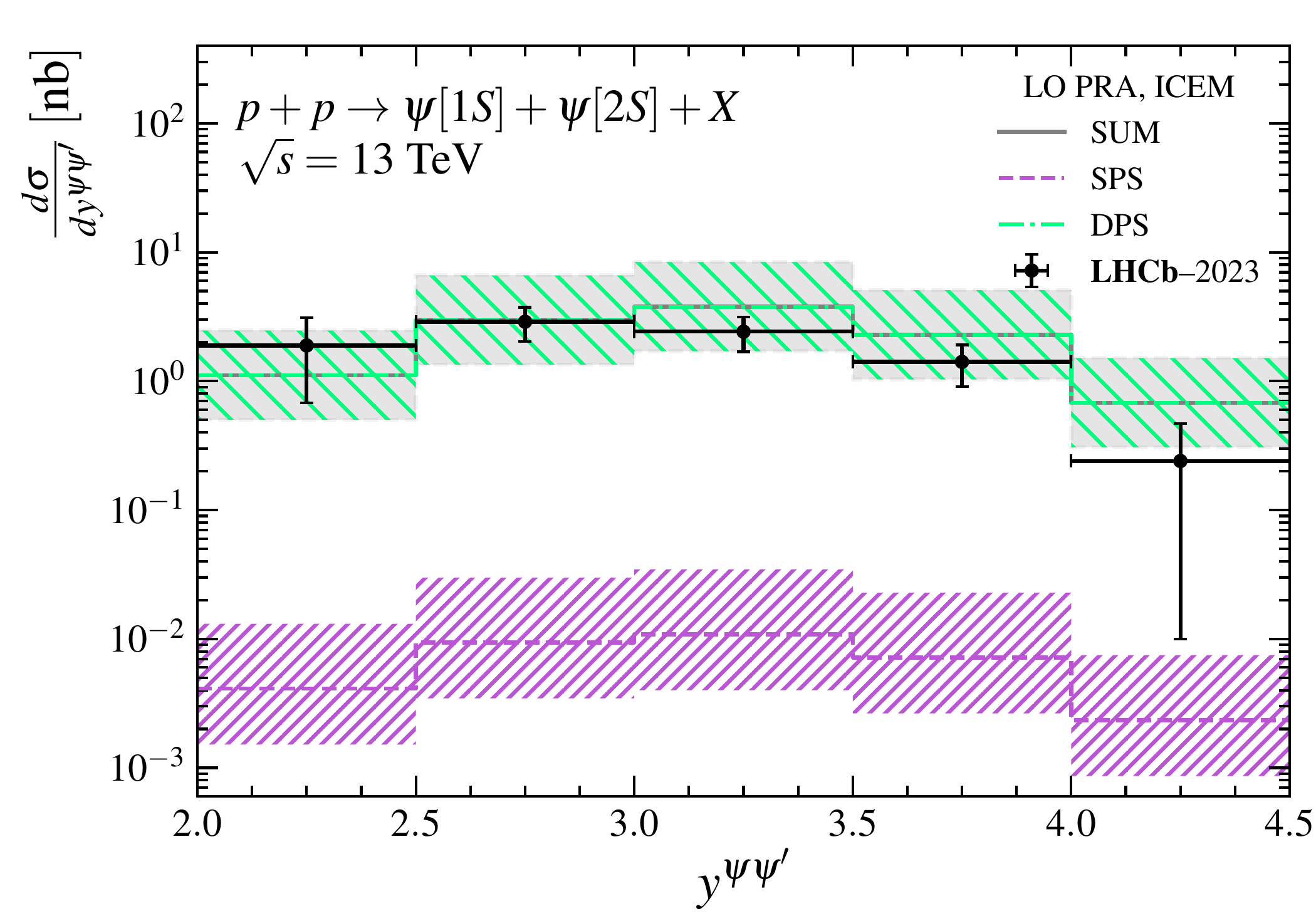}
\\
\includegraphics[scale=0.15]{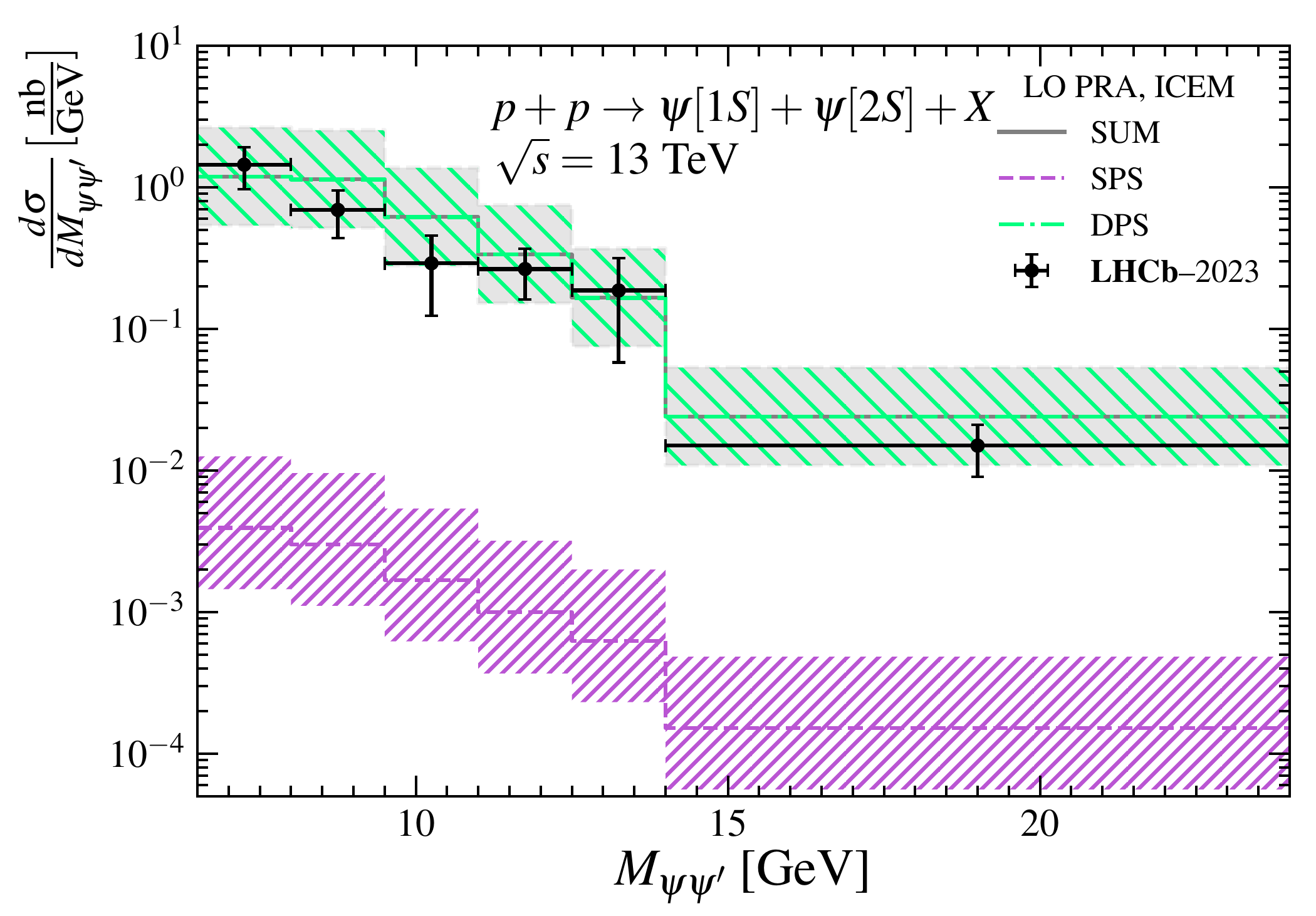}
\caption{\centering
Differential cross sections for pair $\psi[1S] + \psi[1S]$ production at
$\sqrt{s} = 13$ TeV on different kinematical variables. Contributions from SPS
and DPS mechanisms are shown separetly.
The data are from LHCb Collaboration~\cite{LHCb:Psi12S}.}\label{fig:3}
\end{figure}

The comparison of our predicted total cross section in the ICEM
using the PRA for the sum of SPS and DPS contribution is
demonstrated in Table~\ref{tab:1}. An agreement with data within the
experimental errors is achieved. The ratio of the SPS to the DPS is
founded in the ICEM using PRA calculation to be $(3.1
^{+10.1}_{-3.1}) \times 10^{-3}$. In Table~\ref{tab:1} are also
presented values of the ratios $\sigma^{\rm SPS +
DPS}_{\psi[1S]\psi[2S]}/\sigma^{\rm SPS + DPS}_{\psi[1S]\psi[1S]}$
and $\sigma^{\rm DPS}_{\psi[1S]\psi[2S]}/\sigma^{\rm
DPS}_{\psi[1S]\psi[1S]}$ measured for the first time at the LHCb and
our predictions. The comparison of the predicted in the ICEM using
the PRA the differential cross sections with the LHCb  data is
demonstrated in Fig.~\ref{fig:3}. Theoretical uncertainties due to
the hard scale variation by factor 2 relative the central value $\mu
= \left( m_T^{\psi[1S]} + m_T^{\psi[2S]} \right) / 2$ are shown as
shaded areas. We see that the DPS mechanism in the ICEM using the
PRA approach describe data well and contribution of the SPS is even
 much smaller than in the case of $\psi[1S]+\psi[1S]$ pair
production.

\section{Conclusions}\label{sec:conc}

The processes of pair $\psi[1S] + \psi[1S]$ and $\psi[1S] +
\psi[2S]$ production in the forward rapidity region within
frameworks of the ICEM using the PRA and taking into account both
contributions from the SPS and the DPS mechanisms were studied. We
obtain a quite satisfactory description for spectra and cross
sections for the LHCb data at the $\sqrt{s}=13$ TeV. We confirm our
previously obtained value for the DPS pocket--formula parameter
$\sigma_{\rm eff} \simeq 11$ mb, which is in good agreement with new
extracted value of the LHCb Collaboration~\cite{LHCb:2Psi2023}.
Instead of estimations performed in the NLO${}^\star$ approximation
of the CPM using the CSM \cite{Lansberg:2013} as well as in the
NRQCD using PRA \cite{He:2019}, we have found the DPS contribution
is absolutely dominant in the pair charmonium production processes
$\psi[1S] + \psi[1S]$ and $\psi[1S] + \psi[2S]$ at the LHCb.

\section{Acknowledgments}
 We are grateful to Maxim Nefedov for a fruitful
discussion on the heavy quarkonium production at high energy. The
work is supported by the Foundation for the Advancement of
Theoretical Physics and Mathematics BASIS, grant No. 24-1-1-16-5,
and by the grant of the Ministry of Science and High Education of
Russian Federation, No. FSSS-2024-0027.

\newpage

\newpage

\newpage
\bibliographystyle{ws-mpla}
\bibliography{references}

\end{document}